\documentclass{article}
\pdfoutput=1

\usepackage[T1]{fontenc}
\usepackage{amsmath,amssymb,latexsym}
\usepackage[english]{babel}
\usepackage{graphicx}
\usepackage[T2A]{fontenc}
\usepackage[utf8]{inputenc}
\usepackage[unicode,colorlinks,linkcolor=blue,citecolor=red]{hyperref}
\usepackage{cite}

\def\qq{\qquad}
\def\nqq{\hspace*{-2em}}

\def\inch{\hspace*{1in}}

\def\lal{&&\nqq {}}
\def\eq{Eq.\,}

\def\beq{\begin{equation}}
\def\eeq{\end{equation}}
\def\bear{\begin{eqnarray}}
\def\bearr{\begin{eqnarray} \lal}
\def\ear{\end{eqnarray}}
\def\earn{\nonumber \end{eqnarray}}
\def\nn{\nonumber\\ {}}

\def\nnn{\nonumber\\ \lal }

\def\yy{\\[5pt] {}}
\def\yyy{\\[5pt] \lal }

\def\dst{\displaystyle}

\def\fracd#1#2{{\dst\frac{#1}{#2}}}

\def\Half{{\fracd{1}{2}}}

\def\e{{\,\rm e}}

\def\diag{\mathop{\rm diag}\nolimits}

\usepackage{color}

\def\rf{\eqref}
\def\kappa{\varkappa}

\newcommand{\vv}{\text{v}}
\newcommand{\n}{\text{n}}
\newcommand{\D}{\text{D}}

\newcommand{\Pl}{\text{Pl}}
 
\newcommand{\eff}{\text{eff}}

\title{Formation and evolution of a 2-brane structure
in multidimensional $f(R)$ gravity}

\author{{Kirill A. Bronnikov}$^{a,b,c}$\thanks{e-mail: kb20@yandex.ru} \and
{Arkady~A.~Popov}$^{d}$\thanks{e-mail: arkady\_popov@mail.ru} \and
{Sergey~G.~Rubin}$^{c,d}$\thanks{e-mail: sergeirubin@list.ru}}

\date{\em \small
$^a$ Rostest, 119361 Ozyornaya ul. 46, Moscow, Russia\\
$^b$ Inst. of Gravitation and Cosmology, RUDN University,  117198, ul. Miklukho-Maklaya 6, Moscow, Russia\\
$^c$ National Research Nuclear University MEPhI (Moscow Engineering Physics Institute), 115409, Kashirskoe shosse 31, Moscow, Russia \\
$^d$ N.~I.~Lobachevsky Institute of Mathematics and Mechanics, Kazan  Federal  University, 420008, Kremlevskaya  street  18,  Kazan,  Russia}

\begin{document}
\maketitle
%\flushbottom
\begin{abstract}
 It has been previously shown that multidimensional $f(R)$ gravity
 {can lead} to a two-brane structure.
 In this paper, we analyze such a model with a spatially flat 4D de 
 Sitter (dS) cosmology {whose Hubble parameter $H$ determines the 
 universal energy scale}. We show that the two-brane metric is 
 nucleated at the highest energies. The distance between the branes 
 grows gradually as the energy decreases, tending to a finite value at 
 zero energy density.
 
 It is stated that the physical parameters such as the 4D Planck mass, 
 the Higgs vacuum expectation value, and vacuum energy density vary 
 with the evolving universal energy scale, even on the classical level. 
 We also show that the Higgs vacuum expectation value is different on 
 different branes.
\end{abstract}

% ===============================
\section{Introduction}
% ===============================

 Multidimensional gravity is a powerful theoretical framework for addressing 
 fundamental problems in modern physics, from explaining known phenomena to 
 obtaining novel results \cite{Abbott:1984ba, Brown:2013fba, Bronnikov:2009zza, Chaichian:2000az}. 
 Over decades, this approach has yielded significant insights into issues 
 such as the hierarchy problem \cite{Gogberashvili:1998vx,1999PhRvL..83.3370R,ArkaniHamed:1998rs}, 
 the small cosmological constant (CC) \cite{Krause:2000uj, Bronnikov:2023lej}, and the dynamics of multidimensional inflation 
 \cite{2002PhRvD..65j5022G,Bronnikov:2009ai,Fabris:2019ecx}. Much of this 
 work assumes static extra dimensions stabilized at high energy scales, with 
 stabilization itself often arising as a purely gravitational effect 
 \cite{Bronnikov:2005iz,2002PhRvD..66d4014G,2003PhRvD..68d4010G,Arbuzov:2021yai}.

 An important theme within this framework is the brane-world scenario,
 where our observable universe is a 3D hypersurface (brane) embedded 
 in a higher-dimensional bulk. This concept, pioneered by Akama \cite{Akama:1982jy} 
 and Rubakov and Shaposhnikov \cite{Rubakov:1983bb}, has evolved to 
 include ``thick'' branes with internal structure, which offer
 promising mechanisms for localizing various fields 
\cite{Bronnikov:2006bu,Chumbes:2011zt,Hashemi_2018,Dzhunushaliev:2019wvv,Bazeia:2022vac,Wan:2020smy}. 
 A key challenge in any brane model is the localization of fields 
 including gravity, gauge, and spinor fields and the recovery of 
 effective 4D physics \cite{Guo:2023mki, Cui:2020fiz}.

 While the seminal two-brane Randall-Sundrum (RS1) model 
 \cite{Csaki:1999mp,Wang:2008zzr} is well known, it requires a 
 postulated fixed inter-brane distance. This has motivated a study of 
 more dynamic models, including those featuring multiple thick branes 
 \cite{Bronnikov:2007kw, Dzhunushaliev:2019wvv}. In our previous work 
 \cite{Popov:2024nax}, we have demonstrated that one-brane metrics are 
 exceptional, while generic solutions in our framework naturally 
 describe two-brane configurations.

 This paper directly rests on the results of \cite{Popov:2024nax,Rubin:2025xoh}. We study multidimensional gravity with 
 actions that are nonlinear functions of the spacetime curvature, an approach 
 that has previously yielded explanation of some cosmological phenomena such 
 as the fine tuning of fundamental constants and the hierarchy between 
 gravitational and electroweak scales  \cite{Bronnikov:2005iz, Bronnikov:2012wsj, Bronnikov:2023lej}.  This model is more economical than
 our previous construction \cite{Bronnikov:2023lej} as it obviates the initial 
 need for a fundamental scalar field. Classical solutions describing the brane 
 structures exhibit singularities at the brane locations. We suppose that 
 these singularities are smoothed by quantum gravity effects at scales of the 
 order of the D-dimensional Planck length ($l_D$). Provided that the brane 
 thickness is significantly larger than $l_D$, the uncertainties in physical 
 parameters integrated over the extra-dimensional volume remain small.

 Our purpose is to investigate, within this framework, how the effective 4D 
 Planck mass $m_4$ and the Higgs vacuum expectation value $v_h$ depend on 
 the energy scale of the Universe, characterized by the Hubble parameter $H$ 
 of a 4D de Sitter space. We explore this dependence across energies ranging 
 from zero to the Planck scale. This effect is distinct from effects induced
 by quantum corrections. It is therefore of significant interest to study 
 the potential observational implications of these effects acting together.

 Our study reveals that the observed CC $\Lambda$ appears to be related in 
 the standard way to the Hubble parameter $H$ and is independent from 
 the initially postulated D-dimensional CC.

 We also study in detail the distribution of the Higgs field in the extra 
 dimensions, in addition to \cite{Rubin:2025xoh}. It is shown that the 
 Higgs vacuum expectation value (VEV) differs between the two branes. 
 Consequently, the fermion masses on the hidden brane do not coincide with 
 their observable values on our brane.

%To achieve this aim, the knowledge of the branes structure is necessary. In the papers \cite{} the brane structure was revealed numerically. Despite the progress, the exact metric around the branes  cannot be recovered by numerical simulations.
%
% {\textbf{План работ}}
%
%Aim of research: The formation and evolution of the brane structure starting from the highest to the lowest energies. 
%
%The same for the Planck mass and the Higgs vacuum average on both branes.
%
%Detalization:
%
%1. Найти асимптотику метрических функций на обеих бранах типа на Рuc. \ref{figd} и на Рис. \ref{figa}. Necessary in general and for correct calculation of the integrals. Arkady's activity can be found in formulas (35) and below.
%
%2. Графики зависимостей $m_4(H), \Lambda(H) (?), v(H)$ от $H$ и влияние их на наблюдения.
%
%3. График расстояния между бранами в зависимости от Н
%
%4. Сверхтяжелые фермионы на второй бране.

% ========================
\section{Basic equations}
% ========================

   We will consider $D=4+n$-dimensional space-times with the metric
\beq          \label{ds0}
		ds^2 =g_{AB}dX^AdX^B 
		= \e^{2\gamma(u)} \Big( d t^2 -\e^{2H t}(dx^2 +dy^2 +dz^2) \Big)  
			-  du^2 - r(u)^2 d\Omega_{n -1}^2, \quad A, B = 0, \ldots, 4+n,
\eeq	
   It is necessary to make clear in advance which 4D metric should be treated as the observed one.
   If we adhered to a Kaluza-Klein paradigm, we would say that the warp factor  $\e^{2\gamma(u)}$
   must be averaged over the whole extra space. In our present brane-world concept, such averaging 
   must be only extended to the brane region. Moreover, if there are two or more branes, the corresponding
   averages, determining length scales on particular branes, are generically different.  

   With this metric, it is possible to show that solutions of interest can be obtained in the framework of 
   pure (extended) gravity, without matter fields, in a theory based on the action
\bear   \label{S}
		S_g &=&    \frac{m_D^{D-2}}{2}\int d^D x \sqrt{|g_D|}  f(R) 
\ear  	  	
   where $g_D = \det (g_{AB})$,  $f(R)$ is some function (to be chosen later) of the D-dimensional 
   scalar  curvature $R$. Variation of \rf{S} with respect to $g^{AB}$ leads to the field equations
\beq
	-\Half \delta_A^B f (R) + \Big[ R^B_A + \nabla_A \nabla^B - \delta^B_A \Box \Big] f_R = 0,  	 \label{EE}
\eeq		  		  		  		  		  		  		
   where $\Box = \nabla_A \nabla^A$, and $f_R = df/dR$.
	
  Non-coinciding equations \eqref{EE} for the unknown metric functions $\gamma(u)$ and $r(u)$ read
\begin{align}              \label{tt}
& {R'}^2 f_{RRR} +\left[R'' +
\left(3 \gamma'  + (n-1) \frac{r'}{r}\right) R' \right]f_{RR}
- \left( \gamma'' +4{\gamma'}^2 + (n-1)\frac{\gamma' r'}{r}
- \frac{3H^2}{\e^{2 \gamma}} \right)  f_{R} 
- \dfrac{f(R)}{2} = 0,
\yy                  \label{uu}
& \left( 4\gamma' + (n-1) \dfrac{r'}{r}  \right)R'\,f_{RR}
- \left( 4 \gamma'' + 4{\gamma'}^2 + (n-1) \dfrac{r''}{r} \right) \, f_R
- \, \dfrac{f(R)}{2} = 0 \,,
\yy                 \label{aa}
& {R'}^2 f_{RRR} \, + \bigg( R'' + \left(4\gamma'  + (n - 2) \dfrac{r'}{r}\right)R'\bigg) \, f_{RR}
-\biggl(\dfrac{r''}{r} + \frac{4\gamma' r'}{r} + (n-2)\dfrac{{r'}^2}{r^2}
- \dfrac{(n-2)}{r^2} \biggr) f_R
 - \, \dfrac{f(R)}{2} = 0,
\end{align}
  where the prime denotes $d/du$. Also, we will use the expression for the Ricci scalar
\beq      \label{Ricci_n}
        R(u)= \frac{12 H^2}{\e^{2 \gamma}} -8\gamma'' -20{\gamma'}^2
        - (n-1) \left( \dfrac{2 r''}{r} + \frac{8\gamma' r'}{r}
        + (n-2) \left(\dfrac{r'}{r}\right)^2\,- \dfrac{(n-2)}{r^2} \right)
\eeq

   The combination $2\times$\eqref{uu}$ - f_R \times $\eqref{Ricci_n}  is the constraint equation
\bearr             \label{cons}
     \biggr(8\gamma' + 2 (n-1) \dfrac{r'}{r} \biggl) R'  f_{RR} + \Biggl(12 {\gamma'}^2
        + (n-1) \biggl(\dfrac{8\gamma' r'}{r} + (n-2)\dfrac{\bigl({r'}^2-1\bigr)}{r^2} \biggr) +R \Biggr)f_R
\nnn \inch
   - \frac{12 H^2}{\e^{- 2\gamma(u)}} f_R\, - f(R)   =  0	.
\ear
   It is important to notice that 
\bearr                    \label{R4Rn}
		R = \e^{-2\gamma}R_4 +R_n, \qq R_4=12 H^2, 
\nnn
		  R_n = -8\gamma'' -20{\gamma'}^2 - (n-1) \left( \dfrac{2 r''}{r} + \frac{8\gamma' r'}{r}
		  		+ (n-2) \left(\dfrac{r'}{r}\right)^2\,- \dfrac{(n-2)}{r^2} \right).
\ear
   Assuming that the extra-dimensional curvature $R_n$ is much larger than $R_4$, one can write
\beq
		f(R) \simeq f(R_n) +f_R(R_n)e^{-2\gamma} R_4 + \frac{f_{RR}(R_n)}{2} e^{-4\gamma}{R_4}^2.
\eeq

\begin{figure}[ht!]
 \centering
\includegraphics[width=0.3\textwidth]{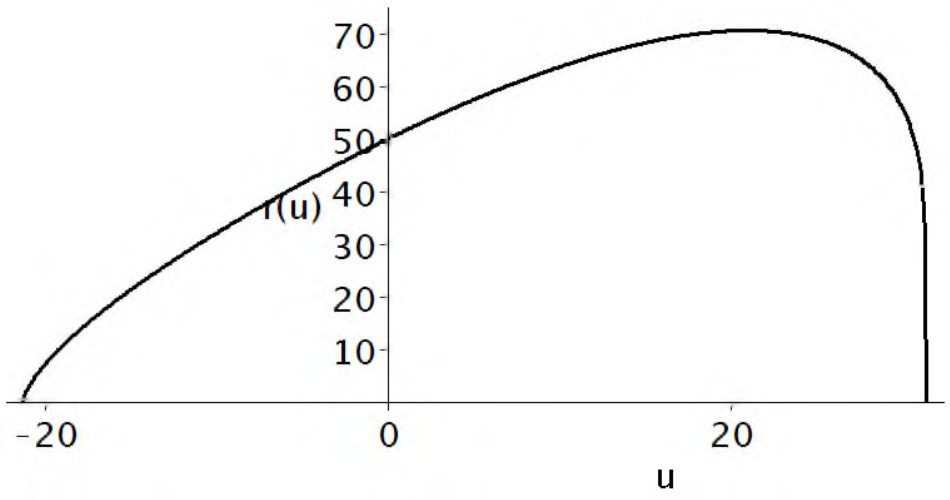}\quad
\includegraphics[width=0.3\textwidth]{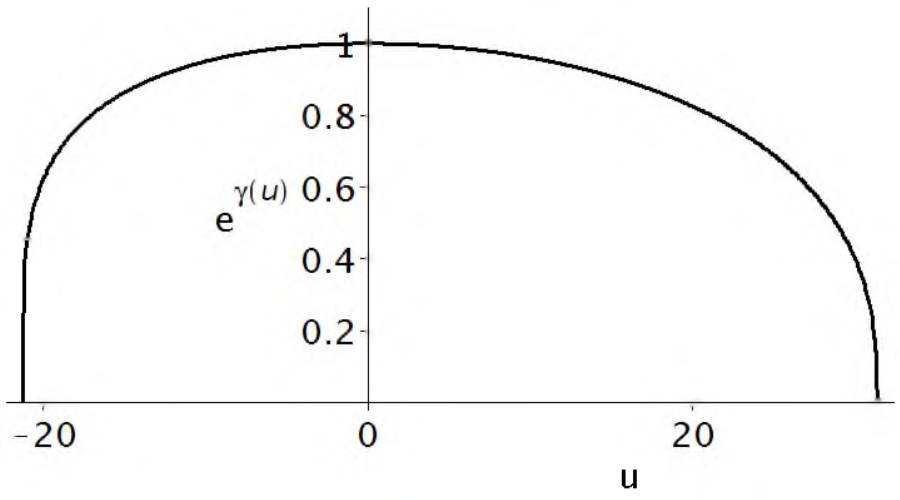}\quad
\includegraphics[width=0.3\textwidth]{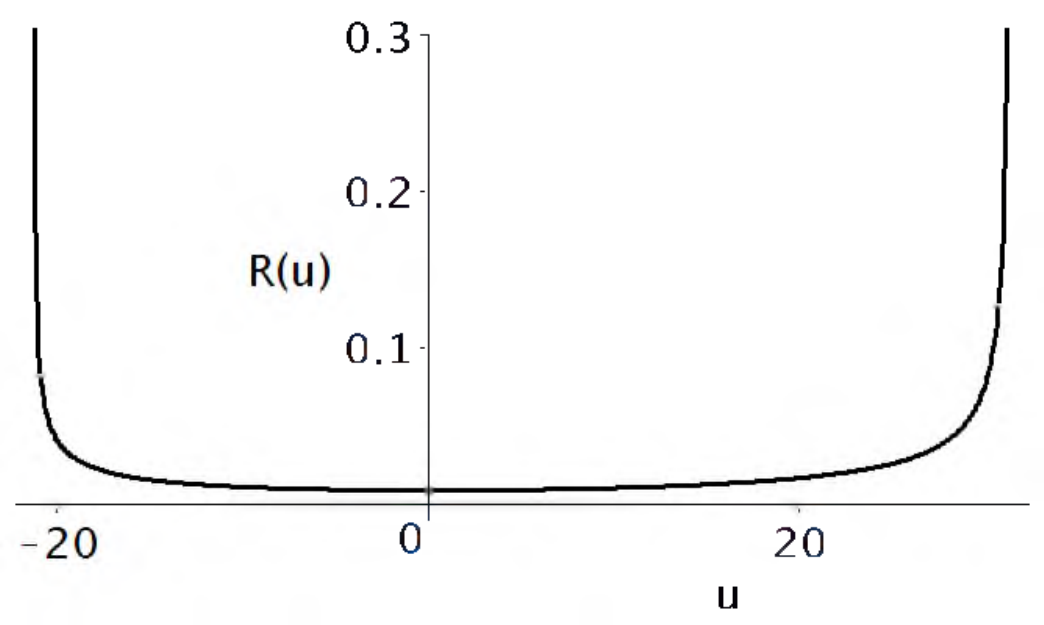}
\caption{\small
	Example of a solution to the equations of motion for $n=2$, $f(R) = 300R^2 +R +0.002$ and $H=0$. 
	Left panel: the extra-dimensional radius $r(u)$, middle: the warp factor $\e^{\gamma(u)}$, 
	right: the Ricci scalar of extra dimensions.  
	The initial conditions: $ r(0)=50, \ \gamma(0) =1, \ r'(0) = 1.6, \ \gamma'(0) = 0, \ 
	R'(0) =-10^{-5}$, and $R(0)$ is determined by \eqref{cons}. 
	}
\label{sol}
\end{figure}

 Fig.\,\ref{sol} presents a solution of interest for us, with the parameters given in the caption.
% ------------------------------------------------------ fig 1  
% ---------------------------------------------------------
  The solution describes a system with two extra dimensions having the shape of a surface of rotation 
  with two singular poles ($r=0$) at which the warp factor $\e^{\gamma(u)}$ tends to zero. The latter 
  circumstance provides the opportunity to treat a close neighborhood of each pole as a brane due 
  their ability to concentrate the fields corresponding to Standard Model particles \cite{Popov:2024nax}.
  The thickness of these branes is defined by matter fields distribution over the extra dimensions. Each brane is considered as independent if the signals between them are forbidden.

% ========================================================
\section{Planck mass variation with the energy scale \label{Planck}}
% ========================================================

 In this paper, we mostly use such units that the $D$-dimensional Planck mass 
 is $m_D=1$, while it is instructive to restore its dimension at the final 
 stage. The specific value of $m_D$ can be found by its relationship with 
 the known 4D Planck mass $m_4$. 

 To this end, one has to take into account the expressions for $R_4$ (the 
 Ricci scalar of our 4D space) and $R_n$ (that of the extra space) given by 
 \rf{R4Rn}, so that the action \eqref{S} takes the form
\bearr               \label{SfR}
S= \frac{m_{D}^{D-2}}{2}  \int_{M_D}  d^{D} X \sqrt{|g_{D}|} \,   f(e^{-2\gamma(u)}R_4(x)+R_n(u))  
\nnn \qq
	 \simeq \frac{m_{D}^{\D-2}}{2}  \int_{M_D}  d^{4} x d^n y \sqrt{|\e^{8\gamma} g_4 \, g_n |} \, [f(R_n(u)) 
	 + e^{-2\gamma(u)}f_R(R_n(u))R_4(x)+ O(R_4^2)] \label{SfR2}
\\  \lal \qq
		\simeq  \frac{{m_4(H)}^2}{2}\int d^4x \sqrt{|g_{4}|} [R_4(x)-2\Lambda_4]\, ,
\ear
 where $g_4$ is the determinant of the matrix
\beq
   \diag(\e^{2 \gamma}, -\e^{2 (\gamma+Ht)}, -\e^{2 (\gamma+Ht)}, -\e^{2 (\gamma+Ht)} ).
\eeq
 The first term in \eqref{SfR2} contributes to the vacuum energy density $\Lambda_4$, 
 see a discussion in \cite{Bronnikov:2023lej}:
\beq             
    -2\Lambda_4 = \frac{ m_D^{D-2} v_{n-1}}{m_4(H)^2}\int_{u_\text{min}}^{u_\text{max}}f(R_{n}) \,
    \e^{4\gamma}\,  r^{n-1} \, du \,,
\eeq
 where $v_{n-1}$ is the volume of an $(n-1)$-dimensional sphere of unit radius
\beq              \label{vn-1}
	{v}_{n-1} \equiv \int_{u_\text{min}}^{u_\text{max}} d^{n-1} y \sqrt{|{g}_{n-1}|} =\dfrac{2\pi^{n/2}}{\,\Gamma(n/2)}. 
\eeq
 The second term in \eqref{SfR2} relates the 4D Planck mass $m_4$ to $m_D$ as follows:
\beq  		\label{m4D}
	m_4(H)^2=m_D^{D-2}v_{n-1}I(H),\qquad 
    I(H)\equiv\int_{u_{\min}}^{u_{\max}} du \e^{2\gamma(u)}\, r(u)^{n-1}f_R(R_n),
\eeq
 The Planck mass is fixed at low energies, $m_4 (H=0)=m_{\Pl}$, and we assume that the relation \eqref{m4D} taken at $H=0$ determines the D-dimensional Planck mass:
\beq      \label{mD}
        m_D^{D-2} = \frac{{m_{\Pl}}^{2}}{v_{n-1}I(H=0)}.
\eeq 
 For estimations, we will keep in mind $m_D\sim 10^{-2}m_{\Pl}$ throughout this 
 paper. More accurate calculation based on \eq (17) and the solution presented in
 Fig.\,1, gives $m_D\sim 3\cdot 10^{-3}m_{\Pl}$.

\begin{figure}[ht!]
 \begin{center}
\includegraphics[width=0.5\linewidth]{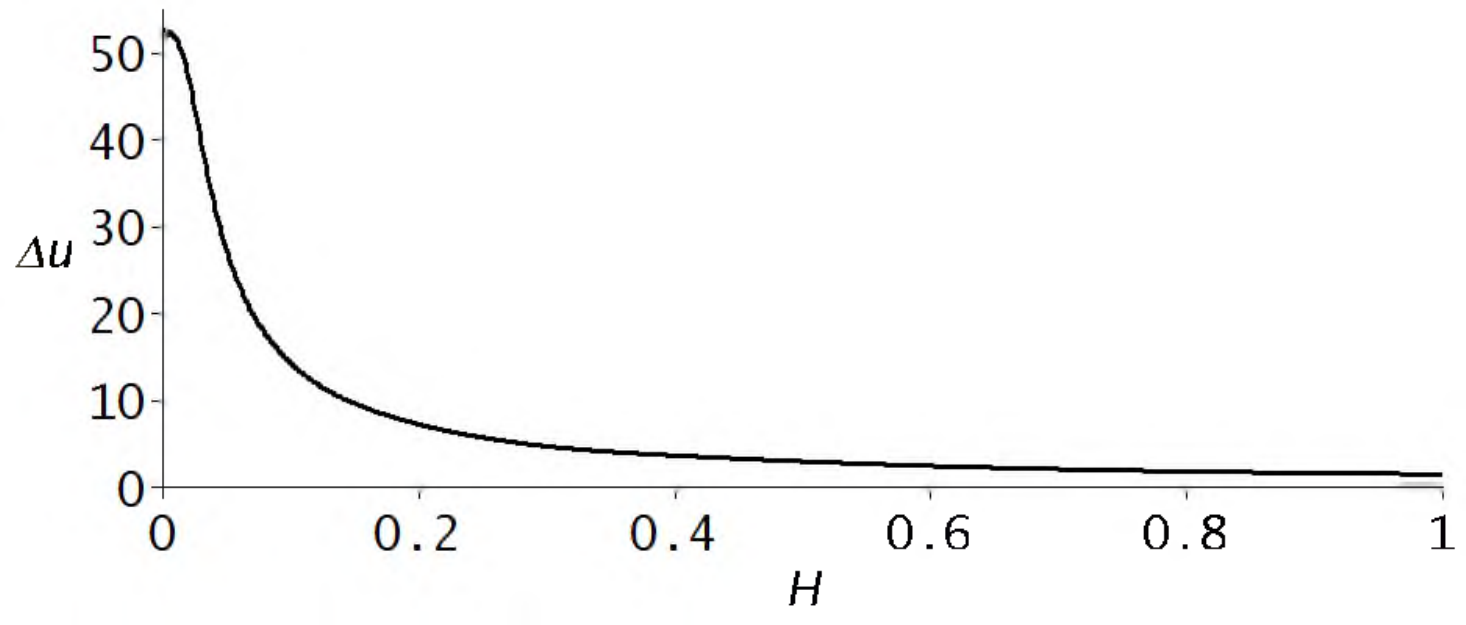}\ 
\includegraphics[width=0.45\linewidth]{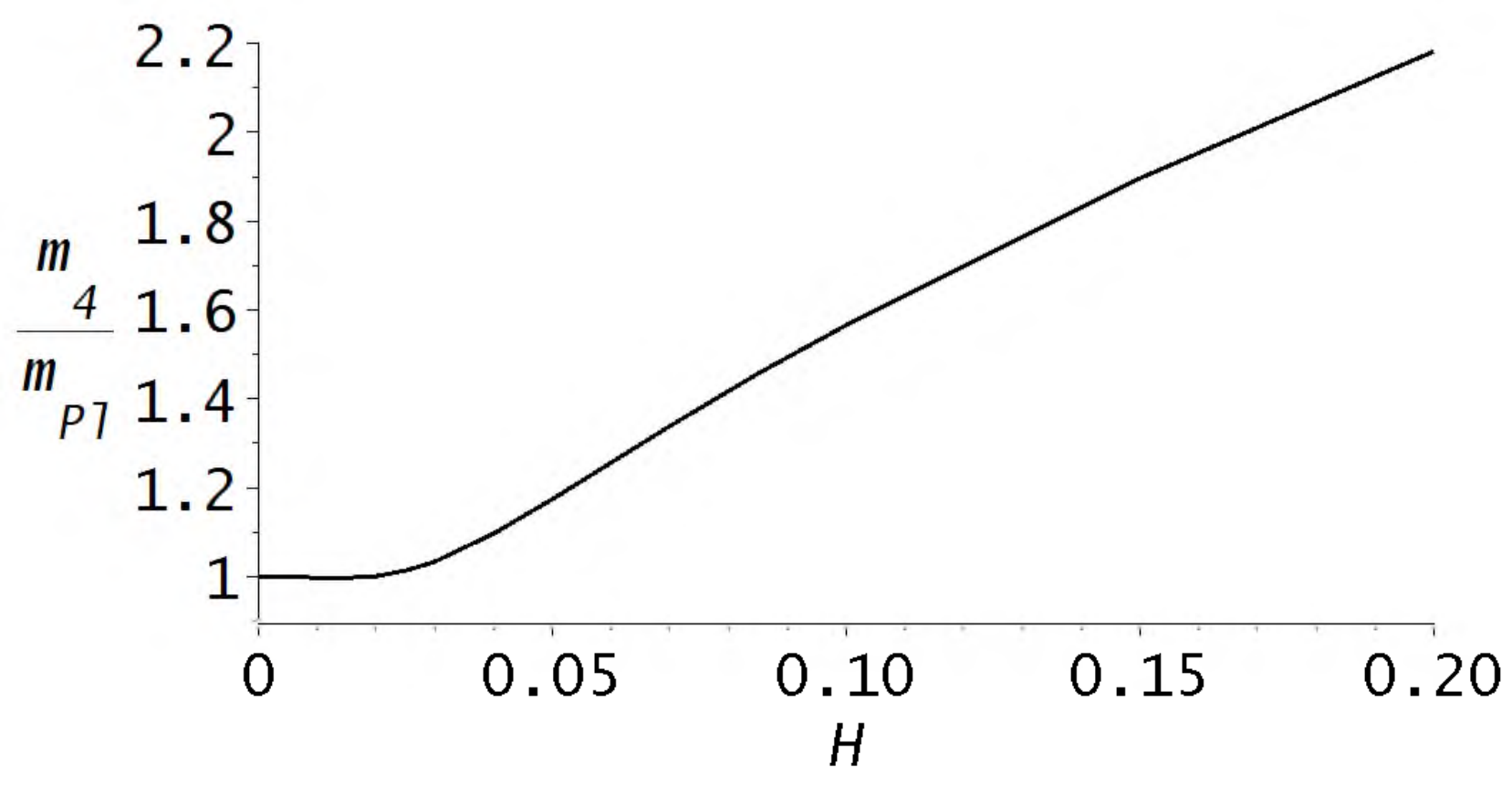}
\end{center}
\caption{\small 
    Left panel: the distance between branes, right panel: variation of the Planck mass $m_4$ with the Hubble parameter. The solution parameters are: $n=2, 
    f(R) = 300R^2 +R +0.002$. The curves are parametrized by the conditions 
    $r(0)=50, \ \gamma(0) =0, \  r'(0) = 1.6, \ \gamma'(0) = 0, \ 
    R'(0) =-10^{-5}$, $R(0)$ is determined by \eqref{cons}.  }
 \label{fige}
\end{figure}

 This means that a quantum regime dominates at $H > 1 =m_D \sim  10^{-2}m_{\Pl}$.
 Knowledge of the metric functions $\gamma(u), r(u)$  and their dependence on $H$ 
 % (see e.g. Figure \ref{metricfig}) 
 permits us to restore the 4D Planck mass $m_4(H)$ for specific values of the 
 Hubble parameter $H$ at an arbitrary energy scale:
\beq                     \label{m4}
        m_4(H)^2={m_{\Pl}}^2\frac{I(H)}{I(H=0)}.
\eeq
 As can be seen from Fig.\,\ref{fige}, right panel, the variation of the 4D Planck 
 mass $m_4(H)$ is quite smooth. Recall that we characterize the cosmological 
 energy scale by the Hubble parameter.

Let us note that the integral in \rf{m4D} encompasses the whole range of 
 the variable $u$ since gravity (unlike matter fields) is assumed to propagate 
 in the whole bulk { between the branes} rather than within the branes.

 The left panel of Fig.\ref{fige} illustrates the two-brane structure formation. 
 The distance $\Delta u$ between the branes is of the order of unity at $H\simeq 1$ 
 (quantum regime) and growths with Hubble parameter decrease. 

% -------------------------------------- fig 2
% -------------------------------------- 

% ==============================================================
\section{Cosmological constant variation with the energy scale}
% ==============================================================

 We assume that the CC is not influenced by any fields, but only $f(R)$ gravity 
 matters. In this case, the CC is found here as a function of the $f(R)$ term acting in $D$ dimensions. 

 The analytical formula for 4D gravity 
\beq                  \label{cH0}
    c_{\eff}(H)  = -2\Lambda_4(H) =  \frac{ m_D^{D-2} v_{n-1}}{m_4(H)^2}\int_{u_{\min}}^{u_{\max}}f(R_n) \,\e^{4\gamma}\,r^{n-1}\,du.
\eeq
 has been obtained in Sec \,\ref{Planck}. 
 As shown in the Appendix, this expression is transformed to the following one:
\bear                 \label{ceff_2}
    c_{\eff} &=& - 6 H^2 + \frac{ m_D^{D-2} v_{n-1}}{m_4(H)^2}\int\limits_{u_{\min}}^{u_{\max}}  
    \Big[ \Big(f_{RR} R' -f_R \gamma'\Big)\,\e^{4\gamma}r^{\n-1}\Big]' \, du 
     + O(H^6).
\ear 
 Let us find the limit of the expression
$ 
\Big[ \Big(f_{RR}\,R'- f_R\,\gamma'\Big)\,\e^{4\gamma} r^{\n-1} \Big]_{u\to u_{0}}\, ,$ where $u_0 =u_{\text{min}} $ {or} $u_{\max}$, 
 assuming that 
\bearr \label{210}
   r(u) \simeq r'(u_0)(u-u_0) +\frac{r''(u_0)}{2}(u-u_0)^2    +\dots\, ,
\yyy    \label{220}
   \e^{\gamma(u)} \simeq  \gamma'(u_0) \e^{\gamma(u_0)}(u-u_0) +\dots \ \ \mbox{or} \ \ 
    \e^{\gamma(u)} \simeq \e^{\gamma(u_0)} + \gamma'(u_0) \e^{\gamma(u_0)}(u-u_0)   +\dots\,\, .
\ear
  It can be shown that this expression tends to zero,
\beq
    \Big[ \Big(f_{RR}\,R'- f_R\,\gamma'\Big)\,
    \e^{4\gamma} r^{\n-1} \Big]_{u\to u_{0}}\, \to 0,
\eeq
 for $f(R)=a R^2 +R +c$, as it takes place in this research.
 At the low energy scale, the Hubble parameter $H$ is small, and the term $H^6$ can be neglected in \eqref{ceff_2} to obtain
\beq    \label{LH}
        c_{\eff}(H/m_D \ll 1)\equiv -2\Lambda_4  = -6\, H^2 \, .
\eeq
 This relation coincides with a well known relation in 4D physics and is valid 
 if $H\ll m_4$. In a more general case of the function $f(R)$, terms proportional 
 to $H^6$ and other nontrivial terms may appear in the expression \eqref{LH}.

 The inflationary dynamic requires a separate study. The Hubble parameter is 
 related to the energy density $\rho$ which is the sum of two terms - the 
 potential $V$ of the inflaton field and the dark energy density proportional to 
 $\Lambda_4$
\beq            \label{6H2_}
    H^2\simeq \frac{8\pi}{3}\frac{\rho}{m_{\Pl}^2}=\frac{8\pi}{3}\frac{V+\frac{\dst m_D^2}{\dst m_{\Pl}^2}\Lambda_4 \frac{\dst m_{\Pl}^2}{\dst 8\pi}}{m_{\Pl}^2}\simeq \frac{8\pi}{3}\frac{V}{m_{\Pl}^2} +H^2\frac{m_D^2}{3m_{\Pl}^2}.
\eeq 
 The factor $m_D^2 / m_{\Pl}^2$ appears at a transition from units $m_D$ 
 to units $m_{\Pl}$. The last equality is a result of the substitution 
 \eqref{LH}. Assuming that $m_D\sim 10^{-2}m_{\Pl}$, the last term is small and can be neglected. Hence, a variation of the CC during inflation is negligible.
 
 In this subsection, we have proved that the well-known relation $H^2=\Lambda_4/3$ can be derived from D-dimensional gravity. For this purpose, we have shown that the boundary term in \eqref{ceff_2} is zero, which is a nontrivial fact. In the standard notations, $c_{\eff}=-2\Lambda_4$, where $\Lambda_4$ is the CC. 

% ================================================================
\section{The Higgs mechanism on different branes}\label{Higgs}
% ================================================================

  As discussed in \cite{Rubin:2025xoh}, a brane other than that inhabited by observers is most likely filled by heavy partners of the Standard Model (SM) particles. According to the SM, the Higgs field is responsible for fermion masses. Therefore, it is of interest to study the Higgs field distribution 
  over the branes. 

% --------------------------------------------------------------
\subsection{Independence of Higgs fluctuations in branes}
% --------------------------------------------------------------

 We begin with a general remark. To describe any system in terms of an action, 
 one must choose a volume of integration. This volume should be chosen so that 
 all fluctuations or particles belonging to the system in question are contained 
 inside it, and the influence of external fields can be neglected. Background fields can be specific to each independent volume.

 In the 2-brane model discussed here, it is natural to choose the integration 
 volume in the action to be bounded by end points denoted as $u_1$ and $u_2$. 
 It is valid under the assumption that field fluctuations describing our world 
 are confined within the interval $u_1$ to $u_2$. A different situation occurs 
 if excitations are localized near a single brane, say, brane-1. In this case, 
 the integration domain shrinks { to the interval $u_1 -  u_1 + \delta$}
 %\blue{and are restricted by some $$, 
 outside which the field fluctuations are negligible. 

 As discussed in \cite{Popov:2024nax}, fermions and various scalar fields are 
 localized near the end points, thereby forming branes. A brief analysis performed there revealed that the Higgs field shares the same property.

\begin{figure}[ht!]
\centering
\includegraphics[width=0.4555\linewidth]{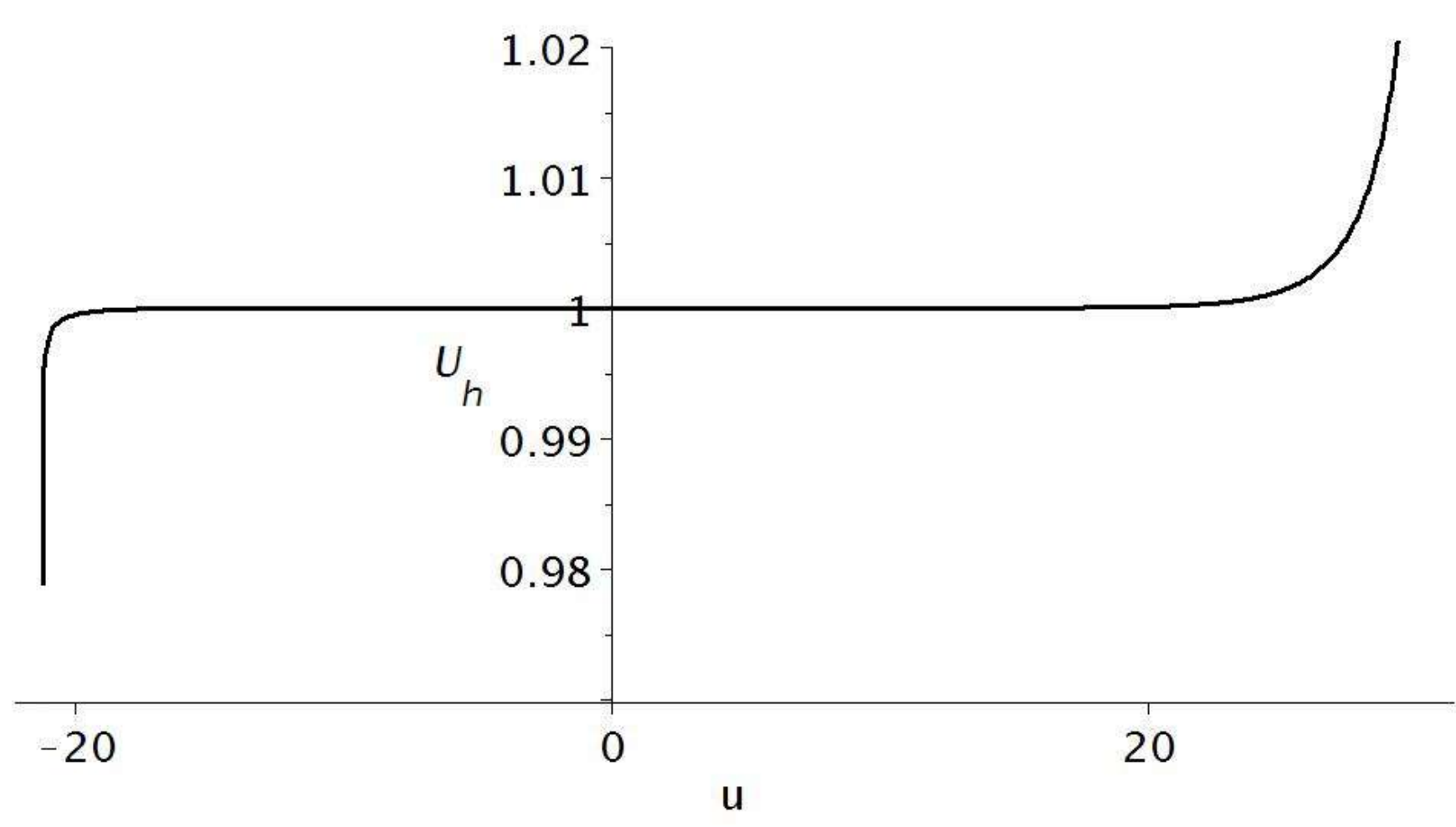}
\caption{\small 
 The Higgs field distribution along the internal coordinate $u$ is specific to each brane. The background metric functions are presented in Fig.\,\ref{sol}.  The parameters are $n=2, f(R) = 300R^2 +R +0.002$, $H=0, \  \nu=0.1, \lambda=0.1$. The curve is parametrized by the conditions 
 $r(0)=50,\ \gamma(0) =0,\ r'(0) = 1.6,\ \gamma'(0) = 0,\ R'(0) =-10^{-5}$, $R(0)$ is determined by \eqref{cons}, $U_h(0)=1,\ U_h'(0)=1 \cdot  10^{-8}$. }
    \label{Higgsfig}
\end{figure}

  The distribution of the Higgs field 
\begin{equation}\label{Hvro}
    H(x,y)=h(x)Y_h(u),\quad h(x) = \frac{1}{\sqrt{2}} 
\begin{pmatrix}
0 \\ v_h +\rho(x)
\end{pmatrix}, \quad v_h \gg\rho(x)
\end{equation} 
 over the extra coordinate $u$ is governed by the scalar function $Y_h(u)$ 
 which is a solution to the classical equation
\begin{equation}        \label{boxUU}
  -\left[\partial^2_u +\left(4\gamma'+(\n-1)\dfrac{r'}{r}\right)\partial_u\right] 
    U_h(u) =\nu U_h(u) - \lambda\, U_h^3(u)\equiv  \frac{d \  V_U(U_h)}{d U_h}. 
\end{equation}
  Here we have made the substitution 
\begin{equation}\label{UYh}
    U_h=v_hY_h
\end{equation}
 The parameter $v_h$ is arbitrary at this stage and will be fixed later by 
 relating it to the observable Higgs vacuum expectation value $v_H$.
 It can be checked numerically that there exist solutions with a slight variation 
 from a constant around the extrema, both minima and maxima, see  
 Fig.\,\ref{Higgsfig}. The {distribution} is stable due to a nontrivial gravitational background. Deviations from the potential minimum at $\sqrt{\nu/\lambda}=1$ are quickly attenuated, a fact to be explained below. 
%
% ---------------------------------  fig 3
% ------------------------------------------

\begin{figure}[ht!]
    \centering
\includegraphics[width=0.45\linewidth]{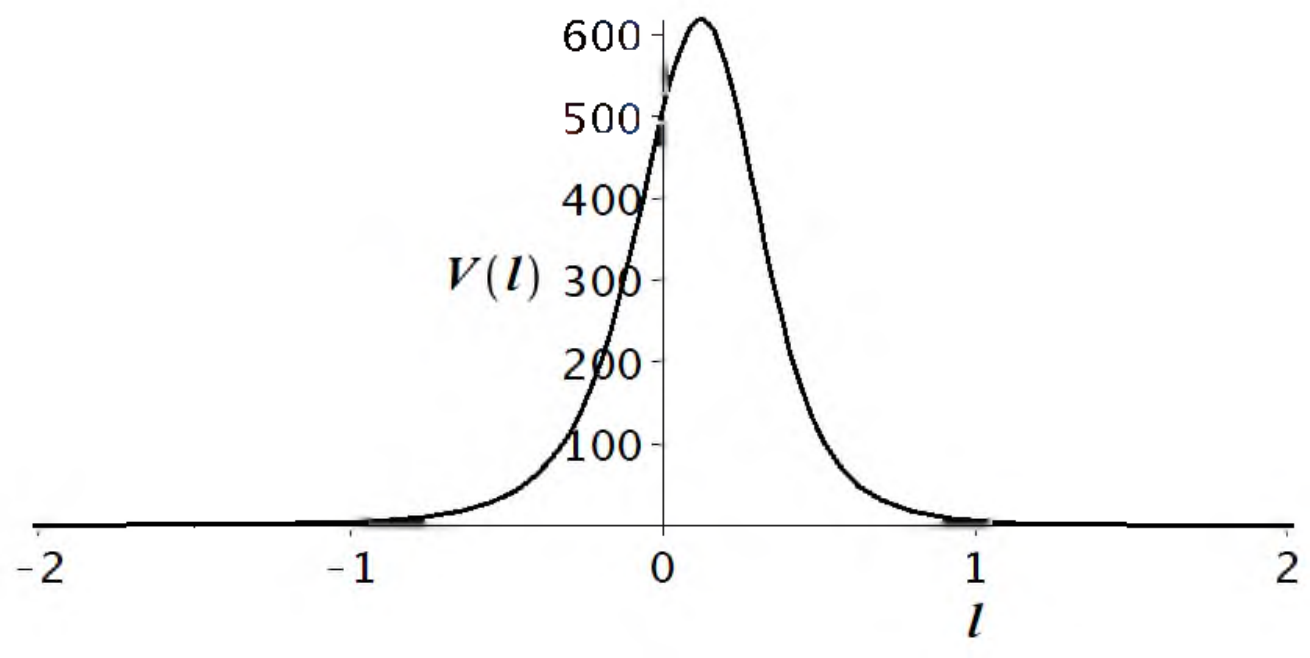} \quad \includegraphics[width=0.35\linewidth]{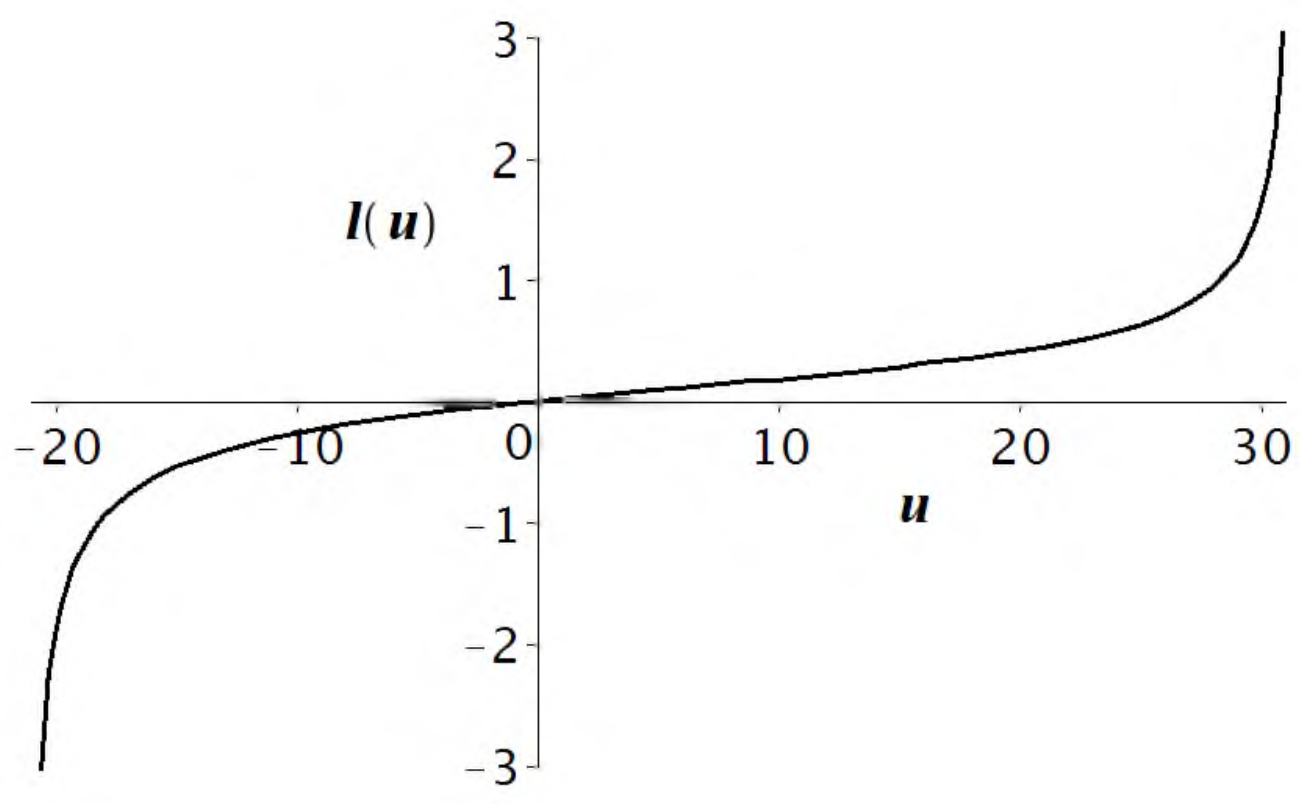} 
\caption{\small
    A barrier between the two branes prevents an exchange of Higgs field fluctuations between them. The parameters are $n=2, f(R) = 300R^2 +R +0.002, H=0$. The curve is parametrized by the conditions $ r(0)=50, \ \gamma(0) =0, \  r'(0) = 1.6, \  \gamma'(0) = 0, \ R'(0) =-10^{-5}$, \ $R(0)$ is determined by \eqref{cons}, $\nu=0.1$.}
    \label{Vl}
\end{figure}

 Here we give a more thorough analysis for the Higgs field considering its 
 small deviation from the potential minimum
\[
        U_{\rm ext} = \sqrt{\nu/\lambda}.
\]        
  Our aim is to show that small deviations located near different branes can be 
  considered independently. To this end, we first exclude the first-order
  derivative in \eq \eqref{boxUU} by moving on to a new variable, $u \to l$
  according to
\beq                 \label{dldu}
        \frac{d l}{d u} = \frac{1}{ r^{n-1} \e^{4 \gamma}  }\, ,
\eeq 
  to obtain
\beq             \label{shred}
    \frac{d^2 U_h}{d l^2} + W(U_h,l) =0, \qquad W(U_h,l) = \Big(\nu U_h(l) - \lambda\, U_h^3(l)\Big)\,   r(l)^{2n-2}\, \e^{8 \gamma(l)}.
\eeq 
  Consider a small variation 
\beq 
        \delta U_h(l)=U_h(l)-U_{\rm ext}\, .
\eeq
  Then
\bear
    W(U_h) &=& \Big[\nu \Big(\delta U_h(l) +U_{ext}\Big) - \lambda\, \Big(\delta U_h(l) +U_{ext}\Big)^3(l)\Big]\,   r(l)^{2n-2}\, \e^{8 \gamma(l)}  
\\
        &=& -V(l)\, \delta U_h(l)+ O(\delta U_h^2)\, ,
\ear
  where
\beq        \label{pot}
        V(l)\equiv 2\nu \,   r(l)^{2n-2}\, \e^{8 \gamma(l)},
\eeq
  which, in the linear approximation, leads to \eq \eqref{shred}:
\begin{equation}               \label{shred2}
        \left[\frac{d^2 }{d l^2} -V(l)\, \right] \delta U_h(l)  =0.
\end{equation}
 This equation is similar to the stationary Schr\"odinger equation with energy $E=0$. The validity of the quasiclassical approximation is confirmed by the presence of a high-magnitude, broad maximum between $u_1$ and $u_2$, as seen 
 in the left panel of Fig.\,\ref{Vl} and is further detailed in the right panel. We conclude that the two types of fluctuations around the potential $V_U$ minima are confined to separate branes and exhibit negligible interaction.

%Suppose that we find solution to \eqref{shred2} with appropriate asymptote $U_h(u\to u_2)$. Then we can solve \eqref{shred} with this asymptote.}
% ------------------------------------- fig 4
% ------------------------------------ 

 Thus we have proved that each brane can be considered as an independent world, 
 the fact that is usually postulated in multi-brane models. Now it is worth 
 showing that the Higgs vacuum average is different in the two branes.
%
%%%As was shown in \cite{Bronnikov:2023lej}, the field distribution over the extra dimensions is specific to each causally separated 3D region below the horizon. Consequently, the field distributions are unique for each universe generated during the de Sitter stage. One of the solutions is shown in Fig. \ref{Higgsfig}. 

%%%The expressions for the Higgs mass and the coupling constant as functionals of the function $U_h(u)$ were obtained in \cite{Bronnikov:2023lej}.\begin{eqnarray}\label{mh}
%&&m^2_h ={v}_{\n-1} \int_{u_{1}}^{u_2} \Bigl(-(\partial_u U_h)^2 +\nu\, U_h^2(u)\Bigr)\e^{4\gamma(u)}r^{\n-1}(u)\, du,
% \\         
%&&\lambda_h = {v}_{\n-1} \int_{u_{1}}^{u_2}\lambda\,U_h^4(u)\,\e^{4\gamma(u)}r^{\n-1}(u)\, du .\label{lh}\end{eqnarray}
%

  As can be seen from Fig.\,\ref{Higgsfig}, the local distribution of the 
  Higgs field varies along the extra coordinate $u$ while its perturbations 
  are locked near one of the branes. The latter is also true for fermions, 
  see  \cite{Popov:2024nax}, so that matter in brane-1 is locked to the 
  interval $u_1 \div u_1+\delta$.

% -------------- We start with the Higgs distributions $H_1(x,u)$ near the brane-1 considering it as independent field  discussed in above. The function $H_1(x,u)$ decreases rapidly far from the endpoint $u_1$ so that the observer on the brane-1 perceives the Higgs field as $H(x,y)=H_1(x,u)=h_1(x)Y^{(1)}_{H}(u)$. The function $H_2(x,u)$ quickly decreases far from the endpoint $u_2$ and an imaginary observer on the brane-2  perceives the Higgs field as $H(x,y)=H_2(x,y)=h_2(x)Y^{(2)}_{H}(u)$.  %Therefore, the Higgs distribution $H$ can be split as $H(x,y)=H_1(x,u)+H_2(x,u)$ with appropriate accuracy. 
% ---------------

  Consider the 4D parameters $m_H$ and $\lambda_H$ related to the Higgs distribution near brane-1. The Higgs field $h(x)$ should be normalized, 
\begin{equation}\label{H0hK}
    H^{(0)}(x)\equiv K_h h(x),
\end{equation} 
 to obtain the standard form of the action, see  \cite{Bronnikov:2023lej}. As was 
 shown there, the parameters $m_H, \lambda_H$ are functionals of the Higgs field 
 distribution near brane-1,
\bear 
    &&m_H^2 \equiv \frac{m_h^2}{K_h},\qquad \lambda_H 
    \equiv \frac{\lambda_h}{K_h^2},       \label{mlH}
\\
    &&    K_h = \frac{{v}_{n-1}}{2v_h^2} \int_{u_1}^{(u_1+\delta)} U_h^2(u)\,\e^{2\gamma(u)}r^{n-1}(u)\,du,     \label{Kh2}
\ear 
  where
\bearr     
    m^2_h =\frac{{v}_{\n-1}}{2v_h^2} \int_{u_1}^{(u_1+\delta)}
    \Bigl(-(\partial_u U_h)^2 +\nu\, U_h^2(u)\Bigr)
    \e^{4\gamma(u)}r^{\n-1}(u)\, du,    \label{m_rho2}
\yyy             \label{lh2}
    \lambda_h = \frac{{v}_{\n-1}}{2v_h^4} \int_{u_1}^{((u_1+\delta)}
    \lambda\,U_h^4(u)\,\e^{4\gamma(u)}r^{\n-1}(u)\, du,
\ear
  where the relation \eqref{UYh} has been taken into account. Here $\nu$ and 
  $\lambda$ are the initial physical parameters of the Higgs field in $D$ 
  dimensions, and ${v}_{n-1}$ is the volume of a unit $(n-1)$-dimensional sphere. Note that integration over the extra dimensions is now carried out across 
  the brane width $\delta$, approximately determined by the width of the 
  matter distribution.

  It is of interest that the physically observed parameters \eqref{mlH} do 
  not depend on the yet unknown value $v_h$ along with the Higgs vacuum average value (VAV),
\begin{equation}\label{vh}
        v_H=\frac{m_H}{\sqrt{2\lambda_H}}.
\end{equation}
   The additional relation 
\begin{equation}
    {v_H}^2=\frac{{v}_{\n-1}}{2} \int_{u_1}^{(u_1+\delta)} U_h^2(u)\,\e^{2\gamma(u)}r^{\n-1}(u)\,du
\end{equation}
  is necessary for self-consistency, as follows from the fact that the VAV 
  of the proto-Higgs field $H(x,y)$, or, more precisely, of $h(x)$ equals 
  to $v_h$ according to  \eqref{Hvro} and the fact that the VAV of the Higgs field $H^{(0)}(x)$ is $v_H$. Also, the relation \eqref{H0hK} and the definition \eqref{Kh2} should be taken into account.

 The above formulas are related to brane-1 where observers are assumed. 
 So the Higgs vacuum average must be equal to the known one, $v_H=246$\,GeV. 
 This can be achieved by choosing appropriate distribution of $U_h$, 
 see \cite{Bronnikov:2023lej}. A strong fine tuning should be made in the expression \eqref{m_rho2} to obtain the observable Higgs mass and its vacuum average. The same formulas with the evident substitution $\int_{u_1}^{u_1+\delta}\to \int^{u_2}_{u_2-\delta}$ can be applied to the second brane where the Higgs distribution is different.  We need not fix 
 the VAV, so that it has an arbitrary value because observers are absent there.

 Therefore, it is of interest to consider the vacuum average $v_H^{(2)}(H=0)$ 
 on brane-2.  Our estimation gives the following Higgs vacuum average value:
\[
    v^{(2)}_H(H=0) \simeq  4\cdot 10^{16} \text{GeV}.
\]
  As a result, for the fields located on brane-2, with the vacuum average $v^{(2)}_H$, the coupling constant $f^{(2)}_e $ and hence the electron 
  mass $m_2\sim f^{(2)}_e v^{(2)}_H$ remain uncertain, varying in a wide 
  range due to the absence of fine tuning. 

% -------------------------------------------------- fig 5
\begin{figure}[ht!]
 \begin{center}
\includegraphics[width=0.6\textwidth]{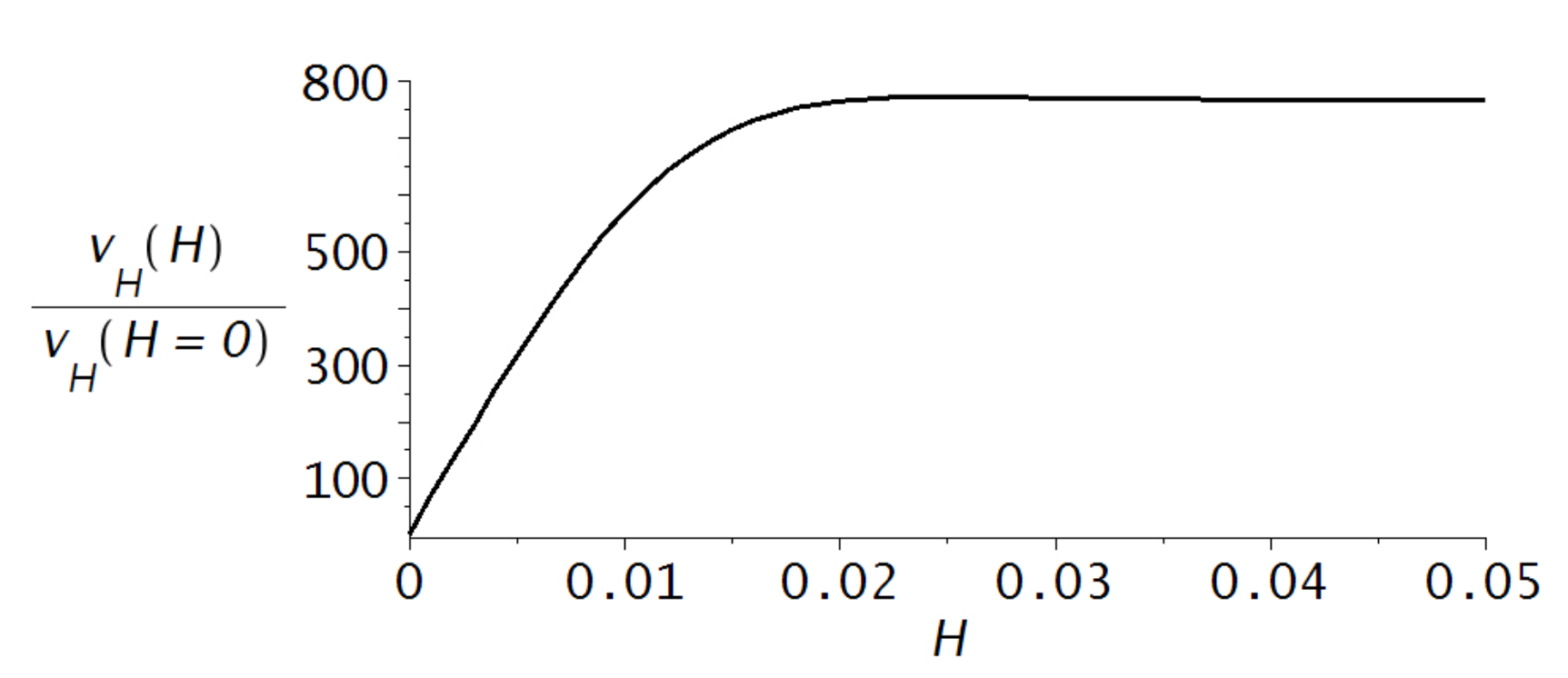}
\end{center} 
\vskip-9mm
\caption{\small
    {The Higgs VEV near brane-1 as a function of $H$.}
    The parameters are $n=2, f(R) = 300R^2 +R +0.002$, $\nu=0.1, \lambda=0.1$. The curve is parametrized by the conditions $ r(0)=50, \ \gamma(0) =0, \  r'(0) = 1.6, \  \gamma'(0) = 0, \ R'(0) =-10^{-5}, \ R(0)$ is determined by \eqref{cons}, $U_h(0)=0.7 , \ U_h'(0)=-1 \cdot  10^{-7}$. }
 \label{figd}
\end{figure}
% ---------------------------------------

  A plot of $v_H(H)$ calculated on brane-1 is presented in Fig.\ref{figd}. 
  As is known, $v_H(H=0)=246$\,GeV. Estimation based on this figure leads to 
  an approximate relation useful for obtaining analytical estimates,
\[
        v_H(H)\simeq (1+12.4\, H)\cdot v_H(H=0).
\]
  This correction is negligibly small at collider energies, but is essential 
  at the inflationary scale.

% ===============================
\section{Conclusion}
% ===============================

  In this paper, we studied multidimensional $f(R)$ gravity with compact 
  extra dimensions, which naturally leads to formation of a dynamical 
  two-brane structure. The key findings of the study are:

1.  Two 4D branes are nucleated at the highest energy scales (near the 
 D-dimensional Planck mass $m_D$). The distance between them is of the order 
 of the D-dimensional Planck scale and grows as the energy scale of the universe, 
 characterized by the Hubble parameter $H$, decreases. This distance approaches 
 a finite value in the low-energy limit ($H \to 0$, see Fig.\,\ref{fige}, 
 left panel).

2. The effective physical parameters, derived from the higher-dimensional 
theory, vary classically with the energy scale. Specifically:

\begin{itemize}
\item 
    The effective 4D Planck mass $m_4$ is a smooth function 
    of $H$, evolving from its known value at $H=0$ to about
    twice that value at high energies (see Fig.\,\ref{fige}, 
    right panel). The effect is not very profound and cannot 
    be observed at present.
\item    
    The effective 4D CC $\Lambda_4$ satisfies the standard 
    4D relation $\Lambda_4 = 3H^2$ at low energies. This 
    relation is shown to hold for an arbitrary value of the 
    initial D-dimensional Lambda term as a nontrivial 
    consequence of the field equations in the energy 
    region where $H\ll m_D$ (see \eqref{LH}). The influence on the inflation rate is negligible.
\item 
    The Higgs vacuum expectation value (VEV) also varies 
    with the energy density. This variation at the inflationary stage is three orders of magnitude higher than the standard value. The same concerns the fermion masses caused by the Higgs mechanism. Nevertheless, the
    particle production rate during inflation seems to be almost the same because the fermion masses remains many orders of magnitude smaller then the inflaton mass. 
\end{itemize}

3. We have shown that the Higgs vacuum expectation value   
  (VEV) differs between the branes. For an observer in 
  our brane (brane-1), it is fine-tuned to the known value 
  of 246 GeV. However, on the second brane (brane-2), the 
  Higgs VEV is much larger, reaching values up to 
  $\sim 10^{16}$\,GeV. This implies that the masses of 
  all Standard Model particles should be enormous on the uninhabited brane-2.

  It is of interest to proceed with multi-brane models, like the one presented in \cite{Arkani-Hamed:1999rvc}.

%In summary, this work presents a framework where the observed 4D space is a brane embedded in a higher-dimensional bulk, with its fundamental constants ($m_4, \lambda_4, v_h$) not being fixed but rather evolving with the cosmological energy scale. 

%The model provides a geometric origin for the hierarchy between the Planck and electroweak scales by segregating them onto different branes, which emerge dynamically from the equations of multidimensional `f(R)` gravity.

\subsection*{Acknowledgements}
The work of SGR and KAB was funded by the Ministry of Science and Higher Education of the Russian Federation, Project "New Phenomena in Particle Physics and the Early Universe" FSWU-2023-0073
and the Kazan Federal University Strategic Academic Leadership Program. The work of AAP was funded by the development program of Volga Region Mathematical Center (agreement No. 075-02-2023-944).
KAB also acknowledges support from Project No. FSSF-2023-0003.

\section*{Appendix}

  In this Appendix, we prove \eq \eqref{ceff_2}. To do that, we manually substitute the expression \eqref{tt} into 
  \eq\eqref{cH0} to obtain
\begin{align}\label{ceff}
 c_{\eff} &  = -2\Lambda_4    =\frac{ m_D^{D-2} v_{n-1}}{m_4(H)^2}\int_{u_{\text{min}}}^{u_{\text{max}}}f(R_{\n}) \, \e^{4\gamma}\,  r^{n-1} \, du  
\nonumber \\
& =\frac{ m_D^{D-2} v_{n-1}}{m_4(H)^2} \int_{u_{\text{min}}}^{u_{\text{max}}} \,
\left\{f\bigl(R_{n}\bigr) +2{R'}^2 f_{RRR}(R) +2\Bigl[R'' +R' \Big(3 \gamma' +(\n-1 ) \dfrac{r'}{r}\Big)  \Bigr] f_{RR}(R)  \right.  
\nonumber \\
& \ \ \ \left. - 2\left( \gamma'' +4{\gamma'}^2 + (\n-1)\frac{\gamma' r'}{r} \right)  f_{R}(R)  
+ \frac{6H^2}{\e^{2\gamma(u)}} f_{R} - f(R) \right\}  \,\e^{4\gamma} r^{n-1} \, du.
 \end{align}
   Part of this expression can be transformed as follows:
\begin{eqnarray} \label{totder}
&& \hspace{-10mm}
\int_{u_{\text{min}}}^{u_{\text{max}}} \,
\left\{ \biggl( 2{R'}^2 f_{RRR} +2\Bigl[R'' +R' \Big(3 \gamma' +(\n-1 ) \dfrac{r'}{r}\Big)  \Bigr] f_{RR} - 2\left( \gamma'' +4{\gamma'}^2  \right. \right.
\nn && \left. \left.
+ (\n-1)\frac{\gamma' r'}{r} \right)  f_{R} \right\} \,\e^{4\gamma} r^{\n-1} \, du
=  \int_{u_{\text{min}}}^{u_{\text{max}}} \Big[ \Big(f_{RR} R' -f_R \gamma' \Big) \,\e^{4\gamma} r^{\n-1}  \Big]' \, du.
\end{eqnarray}
 The Tailor decomposition of the other part of \eqref{ceff} for small $R_4 = 12 H^2 \ll R_\n$ gives:
\bear
&&\int \left[ f(R_n) + \frac{6H^2}{\e^{2\gamma(u)}} f_{R} - f(R)  \right] \,\e^{4\gamma} r^{\n-1} \, du \nn  
&&=\int
\left[f(R_\n) +\frac{6H^2}{\e^{2\gamma(u)}} \left( f_{R}(R_n) +\frac{R_4}{\e^{2\gamma(u)}} f_{RR}(R_n)  \right) - \Big( f(R_n)   \right. 
\nn && \hskip9mm  \left.  
+\frac{R_4}{\e^{2\gamma(u)}} f_{R}(R_n) + \frac{R_4^2}{2\e^{4\gamma(u)}} f_{RR}(R_n) \Big) +O\left( \frac{R_4^3}{\e^{6\gamma(u)}} f_{RRR}(R_n) \right) \right]  \,\e^{4\gamma} r^{\n-1} \, du
 \nn && 
= \int
\left[ \frac{(6H^2 -R_4)}{\e^{2\gamma}} f_R(R_n) 
+\frac{(12H^2 -R_4)R_4}{2\e^{4\gamma}} f_{RR}(R_n)\right. 
\nn && \hskip9mm   \left.
+O\left( \frac{H^6}{\e^{6\gamma(u)}} f_{RRR}(R_n) \right) \right]  \,\e^{4\gamma} r^{\n-1} \, du. 
\ear
Since $m_4(H)^2$ is defined by the expression \eqref{m4D}, the expression \eqref{ceff} is reduced to 
\beq \label{6H2}
c_{\eff} = -2\Lambda_4    = -6 H^2 + O(H^6) + \frac{ m_D^{D-2} v_{n-1}}{m_4(H)^2}\int\limits_{u_{\text{min}}}^{u_{\text{max}}}  \Big[ \Big(f_{RR} R' -f_R \gamma' \Big) \,\e^{4\gamma} r^{\n-1}  \Big]' \, du. 
\eeq
  Thus $c_{\eff}$ is presented in the form of \eqref{ceff_2}.

\bibliographystyle{unsrturl}
\bibliography{Ru-Article_7.bib}

\begin{thebibliography}{10}

\bibitem{Abbott:1984ba}
Richard~B. Abbott, Stephen~M. Barr, and Stephen~D. Ellis.
\newblock {Kaluza-Klein Cosmologies and Inflation}.
\newblock {\em Phys. Rev.}, D30:720, 1984.
\newblock \href {https://doi.org/10.1103/PhysRevD.30.720} {\path{doi:10.1103/PhysRevD.30.720}}.

\bibitem{Brown:2013fba}
Adam~R. Brown, Alex Dahlen, and Ali Masoumi.
\newblock {Compactifying de Sitter space naturally selects a small cosmological constant}.
\newblock {\em Phys. Rev.}, D90(12):124048, 2014.
\newblock \href {https://arxiv.org/abs/1311.2586} {\path{arXiv:1311.2586}}, \href {https://doi.org/10.1103/PhysRevD.90.124048} {\path{doi:10.1103/PhysRevD.90.124048}}.

\bibitem{Bronnikov:2009zza}
K.A. Bronnikov, S.G. Rubin, and I.V. Svadkovsky.
\newblock {High-order multidimensional gravity and inflation}.
\newblock {\em Grav. \& Cosm.}, 15:32--33, 2009.
\newblock \href {https://doi.org/10.1134/S0202289309010083} {\path{doi:10.1134/S0202289309010083}}.

\bibitem{Chaichian:2000az}
Masud Chaichian and Archil~B. Kobakhidze.
\newblock {Mass hierarchy and localization of gravity in extra time}.
\newblock {\em Phys. Lett.}, B488:117--122, 2000.
\newblock \href {https://arxiv.org/abs/hep-th/0003269} {\path{arXiv:hep-th/0003269}}, \href {https://doi.org/10.1016/S0370-2693(00)00874-1} {\path{doi:10.1016/S0370-2693(00)00874-1}}.

\bibitem{Gogberashvili:1998vx}
Merab Gogberashvili.
\newblock {Hierarchy problem in the shell universe model}.
\newblock {\em Int. J. Mod. Phys. D}, 11:1635--1638, 2002.
\newblock \href {https://arxiv.org/abs/hep-ph/9812296} {\path{arXiv:hep-ph/9812296}}, \href {https://doi.org/10.1142/S0218271802002992} {\path{doi:10.1142/S0218271802002992}}.

\bibitem{1999PhRvL..83.3370R}
L.~{Randall} and R.~{Sundrum}.
\newblock {Large Mass Hierarchy from a Small Extra Dimension}.
\newblock {\em Phys. Rev. Lett.}, 83:3370--3373, October 1999.
\newblock \href {https://arxiv.org/abs/arXiv:hep-ph/9905221} {\path{arXiv:arXiv:hep-ph/9905221}}, \href {https://doi.org/10.1103/PhysRevLett.83.3370} {\path{doi:10.1103/PhysRevLett.83.3370}}.

\bibitem{ArkaniHamed:1998rs}
Nima Arkani-Hamed, Savas Dimopoulos, and G.~R. Dvali.
\newblock {The Hierarchy problem and new dimensions at a millimeter}.
\newblock {\em Phys. Lett.}, B429:263--272, 1998.
\newblock \href {https://arxiv.org/abs/hep-ph/9803315} {\path{arXiv:hep-ph/9803315}}, \href {https://doi.org/10.1016/S0370-2693(98)00466-3} {\path{doi:10.1016/S0370-2693(98)00466-3}}.

\bibitem{Krause:2000uj}
Axel Krause.
\newblock {A Small cosmological constant and back reaction of nonfinetuned parameters}.
\newblock {\em J. High Energ. Phys.}, 09:016, 2003.
\newblock \href {https://arxiv.org/abs/hep-th/0007233} {\path{arXiv:hep-th/0007233}}, \href {https://doi.org/10.1088/1126-6708/2003/09/016} {\path{doi:10.1088/1126-6708/2003/09/016}}.

\bibitem{Bronnikov:2023lej}
Kirill~A. Bronnikov, Arkady~A. Popov, and Sergey~G. Rubin.
\newblock {Multi-scale hierarchy from multidimensional gravity}.
\newblock {\em Phys. Dark Univ.}, 42:101378, 2023.
\newblock \href {https://arxiv.org/abs/2307.03005} {\path{arXiv:2307.03005}}, \href {https://doi.org/10.1016/j.dark.2023.101378} {\path{doi:10.1016/j.dark.2023.101378}}.

\bibitem{2002PhRvD..65j5022G}
A.~M. {Green} and A.~{Mazumdar}.
\newblock {Dynamics of a large extra dimension inspired hybrid inflation model}.
\newblock {\em Phys. Rev.~{\bf{D}}}, 65(10):105022, May 2002.
\newblock \href {https://arxiv.org/abs/hep-ph/0201209} {\path{arXiv:hep-ph/0201209}}, \href {https://doi.org/10.1103/PhysRevD.65.105022} {\path{doi:10.1103/PhysRevD.65.105022}}.

\bibitem{Bronnikov:2009ai}
K.A. Bronnikov, S.G. Rubin, and I.V. Svadkovsky.
\newblock {Multidimensional world, inflation and modern acceleration}.
\newblock {\em Phys.\ Rev.\ D}, 81:084010, 2010.
\newblock \href {https://arxiv.org/abs/0912.4862} {\path{arXiv:0912.4862}}, \href {https://doi.org/10.1103/PhysRevD.81.084010} {\path{doi:10.1103/PhysRevD.81.084010}}.

\bibitem{Fabris:2019ecx}
J\'ulio~C. Fabris, Arkady~A. Popov, and Sergey~G. Rubin.
\newblock {Multidimensional gravity with higher derivatives and inflation}.
\newblock {\em Phys. Lett. B}, 806:135458, 2020.
\newblock \href {https://arxiv.org/abs/1911.03695} {\path{arXiv:1911.03695}}, \href {https://doi.org/10.1016/j.physletb.2020.135458} {\path{doi:10.1016/j.physletb.2020.135458}}.

\bibitem{Bronnikov:2005iz}
K.~A. Bronnikov and S.~G. Rubin.
\newblock {Self-stabilization of extra dimensions}.
\newblock {\em Phys. Rev.}, D73:124019, 2006.
\newblock \href {https://arxiv.org/abs/gr-qc/0510107} {\path{arXiv:gr-qc/0510107}}, \href {https://doi.org/10.1103/PhysRevD.73.124019} {\path{doi:10.1103/PhysRevD.73.124019}}.

\bibitem{2002PhRvD..66d4014G}
U.~{G{\"u}nther}, P.~{Moniz}, and A.~{Zhuk}.
\newblock {Asymptotical AdS space from nonlinear gravitational models with stabilized extra dimensions}.
\newblock {\em Phys. Rev.~{\bf{D}}}, 66(4):044014, August 2002.
\newblock \href {https://arxiv.org/abs/hep-th/0205148} {\path{arXiv:hep-th/0205148}}, \href {https://doi.org/10.1103/PhysRevD.66.044014} {\path{doi:10.1103/PhysRevD.66.044014}}.

\bibitem{2003PhRvD..68d4010G}
U.~{G{\"u}nther}, P.~{Moniz}, and A.~{Zhuk}.
\newblock {Nonlinear multidimensional cosmological models with form fields: Stabilization of extra dimensions and the cosmological constant problem}.
\newblock {\em prd}, 68(4):044010, August 2003.
\newblock \href {https://arxiv.org/abs/hep-th/0303023} {\path{arXiv:hep-th/0303023}}, \href {https://doi.org/10.1103/PhysRevD.68.044010} {\path{doi:10.1103/PhysRevD.68.044010}}.

\bibitem{Arbuzov:2021yai}
A.~Arbuzov, B.~Latosh, and A.~Nikitenko.
\newblock {Effective potential of scalar-tensor gravity with quartic self-interaction of scalar field}.
\newblock {\em Class. Quant. Grav.}, 39(5):055003, 2022.
\newblock \href {https://arxiv.org/abs/2109.09797} {\path{arXiv:2109.09797}}, \href {https://doi.org/10.1088/1361-6382/ac4827} {\path{doi:10.1088/1361-6382/ac4827}}.

\bibitem{Akama:1982jy}
K.~Akama.
\newblock {An Early Proposal of 'Brane World'}.
\newblock {\em Lect. Notes Phys.}, 176:267--271, 1982.
\newblock \href {https://arxiv.org/abs/hep-th/0001113} {\path{arXiv:hep-th/0001113}}.

\bibitem{Rubakov:1983bb}
V.~A. Rubakov and M.~E. Shaposhnikov.
\newblock {Do We live inside a domain wall?}
\newblock {\em Phys. Lett.}, 125B:136--138, 1983.
\newblock \href {https://doi.org/10.1016/0370-2693(83)91253-4} {\path{doi:10.1016/0370-2693(83)91253-4}}.

\bibitem{Bronnikov:2006bu}
K.A. Bronnikov, B.E. Meierovich, and S.T. Abdyrakhmanov.
\newblock {Global topological defects in extra dimensions and the brane world concept}.
\newblock {\em Grav.\ Cosmol.}, 12:106--110, 2006.

\bibitem{Chumbes:2011zt}
A.~E.~R. Chumbes, J.~M. Hoff~da Silva, and M.~B. Hott.
\newblock {A model to localize gauge and tensor fields on thick branes}.
\newblock {\em Phys. Rev. D}, 85:085003, 2012.
\newblock \href {https://arxiv.org/abs/1108.3821} {\path{arXiv:1108.3821}}, \href {https://doi.org/10.1103/PhysRevD.85.085003} {\path{doi:10.1103/PhysRevD.85.085003}}.

\bibitem{Hashemi_2018}
S.~Sedigheh Hashemi and Nematollah Riazi.
\newblock Vacuum f(r) thick brane solution with a modified gaussian warp function.
\newblock {\em Annals of Physics}, 399:137--148, 12 2018.
\newblock URL: \url{http://dx.doi.org/10.1016/j.aop.2018.10.010}, \href {https://doi.org/10.1016/j.aop.2018.10.010} {\path{doi:10.1016/j.aop.2018.10.010}}.

\bibitem{Dzhunushaliev:2019wvv}
V.~Dzhunushaliev, V.~Folomeev, G.~Nurtayeva, and S.~D. Odintsov.
\newblock {Thick branes in higher-dimensional $f(R)$ gravity}.
\newblock {\em Int. J. Geom. Meth. Mod. Phys.}, 17(03):2050036, 2020.
\newblock \href {https://arxiv.org/abs/1908.01312} {\path{arXiv:1908.01312}}, \href {https://doi.org/10.1142/S021988782050036X} {\path{doi:10.1142/S021988782050036X}}.

\bibitem{Bazeia:2022vac}
D.~Bazeia and A.~S. Lob\~ao.
\newblock {Mechanism to control the internal structure of thick brane}.
\newblock {\em Eur. Phys. J. C}, 82(7):579, 2022.
\newblock \href {https://arxiv.org/abs/2206.10794} {\path{arXiv:2206.10794}}, \href {https://doi.org/10.1140/epjc/s10052-022-10546-z} {\path{doi:10.1140/epjc/s10052-022-10546-z}}.

\bibitem{Wan:2020smy}
Jun-Jie Wan, Zheng-Quan Cui, Wen-Bin Feng, and Yu-Xiao Liu.
\newblock {Smooth braneworld in $6$-dimensional asymptotically AdS spacetime}.
\newblock {\em JHEP}, 05:017, 2021.
\newblock \href {https://arxiv.org/abs/2010.05016} {\path{arXiv:2010.05016}}, \href {https://doi.org/10.1007/JHEP05(2021)017} {\path{doi:10.1007/JHEP05(2021)017}}.

\bibitem{Guo:2023mki}
Heng Guo, Yong-Tao Lu, Cai-Ling Wang, and Yue Sun.
\newblock {Localization of scalar field on the brane-world by coupling with gravity}.
\newblock {\em JHEP}, 06:114, 2024.
\newblock \href {https://arxiv.org/abs/2310.01451} {\path{arXiv:2310.01451}}, \href {https://doi.org/10.1007/JHEP06(2024)114} {\path{doi:10.1007/JHEP06(2024)114}}.

\bibitem{Cui:2020fiz}
Zheng-Quan Cui, Zi-Chao Lin, Jun-Jie Wan, Yu-Xiao Liu, and Li~Zhao.
\newblock {Tensor Perturbations and Thick Branes in Higher-dimensional $f(R)$ Gravity}.
\newblock {\em JHEP}, 12:130, 2020.
\newblock \href {https://arxiv.org/abs/2009.00512} {\path{arXiv:2009.00512}}, \href {https://doi.org/10.1007/JHEP12(2020)130} {\path{doi:10.1007/JHEP12(2020)130}}.

\bibitem{Csaki:1999mp}
Csaba Csaki, Michael Graesser, Lisa Randall, and John Terning.
\newblock {Cosmology of brane models with radion stabilization}.
\newblock {\em Phys. Rev. D}, 62:045015, 2000.
\newblock \href {https://arxiv.org/abs/hep-ph/9911406} {\path{arXiv:hep-ph/9911406}}, \href {https://doi.org/10.1103/PhysRevD.62.045015} {\path{doi:10.1103/PhysRevD.62.045015}}.

\bibitem{Wang:2008zzr}
Anzhong Wang, Rong-Gen Cai, and N.~O. Santos.
\newblock {Two 3-branes in Randall-Sundrum setup and current acceleration of the universe}.
\newblock {\em Nucl. Phys. B}, 797:395--430, 2008.
\newblock \href {https://arxiv.org/abs/astro-ph/0607371} {\path{arXiv:astro-ph/0607371}}, \href {https://doi.org/10.1016/j.nuclphysb.2007.11.009} {\path{doi:10.1016/j.nuclphysb.2007.11.009}}.

\bibitem{Bronnikov:2007kw}
K.~A. Bronnikov and B.~E. Meierovich.
\newblock {Global strings in extra dimensions: A Full map of solutions, matter trapping and the hierarchy problem}.
\newblock {\em J. Exp. Theor. Phys.}, 106:247--264, 2008.
\newblock \href {https://arxiv.org/abs/0708.3439} {\path{arXiv:0708.3439}}, \href {https://doi.org/10.1007/s11447-008-2005-0} {\path{doi:10.1007/s11447-008-2005-0}}.

\bibitem{Popov:2024nax}
Arkadiy~A. Popov and Sergey~G. Rubin.
\newblock {Spontaneous Brane Formation}.
\newblock {\em Symmetry}, 17(2):252, 2025.
\newblock \href {https://arxiv.org/abs/2408.14692} {\path{arXiv:2408.14692}}, \href {https://doi.org/10.3390/sym17020252} {\path{doi:10.3390/sym17020252}}.

\bibitem{Rubin:2025xoh}
Sergey~G. Rubin.
\newblock {Astrophysical signals from the neighboring brane}.
\newblock 2 2025.
\newblock \href {https://arxiv.org/abs/2502.13761} {\path{arXiv:2502.13761}}.

\bibitem{Bronnikov:2012wsj}
Kirill~A. Bronnikov and Sergey~G. Rubin.
\newblock {\em {Black Holes, Cosmology and Extra Dimensions}}.
\newblock WSP, 2012.
\newblock \href {https://doi.org/10.1142/12186} {\path{doi:10.1142/12186}}.

\bibitem{Arkani-Hamed:1999rvc}
Nima Arkani-Hamed, Savas Dimopoulos, G.~R. Dvali, and Nemanja Kaloper.
\newblock {Many fold universe}.
\newblock {\em JHEP}, 12:010, 2000.
\newblock \href {https://arxiv.org/abs/hep-ph/9911386} {\path{arXiv:hep-ph/9911386}}, \href {https://doi.org/10.1088/1126-6708/2000/12/010} {\path{doi:10.1088/1126-6708/2000/12/010}}.

\end{thebibliography}

%\end{thebibliography}
\end{document}